\documentclass[a4paper, 10pt, conference]{IEEEtran}
\IEEEoverridecommandlockouts
\usepackage{cite}
\usepackage{amsmath,amssymb,amsfonts}
\usepackage{algorithmic}
\usepackage{graphicx}
\usepackage{textcomp}
\usepackage{xcolor}
\usepackage{comment}
\usepackage[flushleft]{threeparttable}
\usepackage{booktabs}
\usepackage{csquotes}
\usepackage{subfig}
\usepackage{tikz}

\def\BibTeX{{\rm B\kern-.05em{\sc i\kern-.025em b}\kern-.08em
    T\kern-.1667em\lower.7ex\hbox{E}\kern-.125emX}}
\begin{document}

\title{The Impact of Social Environment and Interaction Focus on User Experience and Social Acceptability of an Augmented Reality Game}

\author{
 \IEEEauthorblockN{Lorenzo Cocchia$^{1,4}$, Maurizio Vergari$^1$, Tanja Koji\'c$^1$, Francesco Vona$^{3,4}$,\\ Sebastian M\"oller$^{1,2}$, Franca Garzotto$^4$, Jan-Niklas Voigt-Antons$^3$}
 \IEEEauthorblockA{$^1$Quality and Usability Lab, TU Berlin, Germany \\  $^2$German Research Center for Artificial Intelligence (DFKI), Berlin, Germany\\ $^3$Immersive Reality Lab, Hamm-Lippstadt University of Applied Sciences, Germany \\ $^4$Department of Electronics, Information and Bio-engineering, Politecnico di Milano, Italy}
}

\maketitle

\begin{abstract}
One of the most promising technologies inside the Extended Reality (XR) spectrum is Augmented Reality. This technology is already in people's pockets regarding Mobile Augmented Reality with their smartphones. The scientific community still needs answers about how humans could and should interact in environments where perceived stimuli are different from fully physical or digital circumstances. Moreover, it is still being determined if people accept these new technologies in different social environments and interaction settings or if some obstacles could exist. This paper explores the impact of the Social Environment and the Focus of social interaction on users while playing a location-based augmented reality game, measuring it with user experience and social acceptance indicators. An empirical study in a within-subject fashion was performed in different social environments and under different settings of social interaction focus with N = 28 participants compiling self-reported questionnaires after playing a Scavenger Hunt in Augmented Reality. The measures from two different Social Environments (Crowded vs. Uncrowded) resulted in statistically relevant mean differences with indicators from the Social Acceptability dimension.
Moreover, the analyses show statistically relevant differences between the variances from different degrees of Social Interaction Focus with Overall Social Presence, Perceived Psychological Engagement, Perceived Attentional Engagement, and Perceived Emotional Contagion. The results suggest that a location-based AR game played in different social environments and settings can influence the user experience's social dimension. Therefore, they should be carefully considered while designing immersive technological experiences in public spaces involving social interactions between players.
\end{abstract}

\newcommand\copyrighttext{%
  \footnotesize \textcopyright 2024 IEEE. Personal use of this material is permitted. Permission from IEEE must be obtained for all other uses, in any current or future media, including reprinting/republishing this material for advertising or promotional purposes, creating new collective works, for resale or redistribution to servers or lists, or reuse of any copyrighted component of this work in other works. DOI and link to original publication will be added as soon as they are available.}
  
\newcommand\copyrightnotice{%
\begin{tikzpicture}[remember picture,overlay,shift={(current page.south)}]
  \node[anchor=south,yshift=10pt] at (0,0) {\fbox{\parbox{\dimexpr\textwidth-\fboxsep-\fboxrule\relax}{\copyrighttext}}};
\end{tikzpicture}%
}

\copyrightnotice

\section{Introduction and Related Work}
Extended Reality (XR) technologies, which include Augmented Reality (AR), Mixed Reality (MR), and Virtual Reality (VR), are currently seeing a big increase in economic interest as this study is being conducted. It's expected that, due to this trend, more people will soon be regularly using AR, VR, or MR applications. Many leading digital companies that greatly influence various industries invest money in XR technologies \cite{kelly2022forbes}. An important issue they are dealing with is how users should interact with one another. It's believed that \enquote{The Metaverse [...] will change the future of online social networking} \cite{cai2022compute}. Therefore, the paper wants to investigate how individuals should use these technologies and how they should interact with each other through XR technologies. Even without considering technology, this subject understanding is already a complex matter. With the expectation that these technologies will become widely available, it's predicted that they will also change how people socialize, especially in public places.

%Tanja moved RQ into objectives sections under RW, as here it is too soon to introduce it, so that I would make both in the Introduction section having RW and objectives (we have been doing it in this form before)

    \subsection{User Experience in AR and Games}

    User Experience (UX) is a measurable aspect of all products and services. Ongoing research in UX spans various areas and industries, making it challenging to identify a unified approach for broad fields like AR. In one study \cite{arifin2018user}, an attempt was made to establish standard UX metrics specifically for the educational sector. Another research \cite{hammady2018museum} concentrated on UX in cultural applications, such as heritage museums. Regarding Mobile AR (MAR), it was noted \cite{irshad2014user} that the methodology for designing high-quality UX still needs to be explored despite an extensive literature review. A framework to enhance UX by fostering emotional connections with MAR was suggested \cite{dirin2018user}. The domain of gaming offers a distinct experience and product category, with dedicated research focusing on the UX quality of games. A study \cite{persada2019user} outlined three dimensions for evaluating UX in games, aiming to improve player interaction: cognitive factors (like learnability and memorability), social factors (such as social influence and behavioral intentions), and technical factors (providing engaging tools or surveys to identify user personas). However, these frameworks serve as guidelines rather than precise methodologies for systematic research.

    \subsection{Social Acceptability}
    Social acceptability is often defined through its absence in the Human-Computer Interaction (HCI) field, as highlighted by one study \cite{koelle2020social}. The APA Dictionary of Psychology describes it as the lack of disapproval. Similarly, another study on wearable technology \cite{kelly2016wear} characterizes it as the absence of negative feedback or judgment from others. Regarding mobile devices, it is described as a balance between the desire to use technology and the social norms of the environment, involving more than just avoiding embarrassment or being polite. It includes a range of factors such as appearance, social status, and cultural norms \cite{rico2010usable}.
    
    \subsection{Social Context}
    The environment around users when they use a product is a crucial factor affecting their overall experience. This context is often considered more significant than previously thought in discussions on UX. Among various definitions of UX, the concept of \enquote{context} frequently appears as a key element. For instance, UX is defined as the outcome of actions motivated by a specific context in one study \cite{makela2001supporting}. Recent research has further explored the relationship between environment and UX \cite{sonderegger2019ux}.
    
        \subsubsection{Public Spaces}
        Research focusing solely on public spaces, including studies on public delivery services, public display applications, and VR experiences in public areas, often seeks to identify design patterns or external factors influencing UX beyond the physical setting \cite{chen2021ai, keskinen2013evaluating, eghbali2019social}. In their analysis, these investigations commonly treat public spaces as a backdrop rather than a variable. An interesting concept is Social Environment, which can be defined by the number of people, their proximity, and possibly their behavior in the environment \cite{vergari2021influence}. When there are no people in such an environment, we refer to it as \enquote{\textbf{\textit{Uncrowded}}}, while if some people are present, it is called \enquote{\textbf{\textit{Crowded}}}.
        
        \subsubsection{Social Interactions}
        Social interaction is defined for games in \cite{zagal2000model} as a purposeful bilateral communication that can be either natural or stimulated by the game, depending on the rules of the game: if rules are encouraging players to interact, then the game has a stimulated type of interaction. Otherwise, it is natural. Furthermore, it is relevant to understand all the possible interactions that could be accomplished while playing a game. Fonseca and their team made a relevant contribution to this field \cite{fonseca2021designing,fonseca2021requirements,fonseca2022design}. The first distinction is having or not having a common goal between players. This is defined as \enquote{\textit{\textbf{Focused}}} if people have common goals, \enquote{\textbf{\textit{Unfocused}}}otherwise.
        
        \subsection{Social Presence}
        Social Presence needs a universally agreed-upon definition within the scientific community since it sometimes needs to be clarified. Despite this, the work in this area often serves as a foundational reference. The distinction between Telepresence, or Spatial Presence, and Social Presence — the former relating to the sensation of \enquote{being there} and the latter to the feeling of \enquote{being together with another} — is made clear. While these definitions may initially not cover all aspects, they often implicitly include the mental constructs of such spaces, which foster a sense of spatial illusion.
        In other words, in the Human-Computer Interaction field, the concept of Presence examines how people perceive and interact with various technologies \cite{biocca2003toward}. This research led to the development of the Networked Minds Social Presence Inventory \cite{NMPI6743}, a validated tool based on self-reports for evaluating Social Presence.

\subsection{Objectives}
Overall, an interesting focus is on things that could affect UX and how people socially interact when they use these applications in public places. This work first aims to determine whether social settings can change the UX and how socially acceptable XR technology is. Also, it looks closely at social interactions to see if having a common goal affects UX and social acceptability. Another key point is to look at Social Presence, which is about how people feel and connect in these environments, and see how it's influenced by the public setting and how people interact. To explore these questions, a study that exploits a mobile AR game of a Scavenger Hunt was designed, leading to the following research questions:

\begin{itemize}
    \item To what extent can User Experience and/or Social Acceptability in a Mobile Augmented Reality game be affected by Social Environments and the Social Interaction Focus?
    \item Can observations of users' Social Presence while playing a Mobile Augmented Reality game under various Social Environments and Social Interaction Focus conditions provide any insights?

\end{itemize}

%rephreased until here by Tanja

\section{Methods}

    \subsection{Participants}
    The experiment included 28 participants (N= 28, 17 female, 11 male). The average age of participants was 22.54 (median = 23, SD = 1.05). Participants' education was self-declared. Fifteen declared to be in the process of obtaining a Bachelor's Degree, 12 of them are Master's Degree or Master in Business Administration (MBA) candidates, while 1 participant declared not to be a scholar of any higher education institute. The Affinity for Technology Interaction \cite{franke2019ati} scores resulted in a mean value of ATI= 4.13 (median = 4.44, SD = 0.85).
    Participants were recruited from the Polytechnic University of Milan and word of mouth.

    \subsection{Test Setup}
    As an experimental setting, a city landmark and public park were chosen. Moreover, a 2x2 factorial within-subject design with randomization of conditions has been set up. The study's data were collected using web-based surveys that included open-ended and multiple-choice items. In total, there were four experimental conditions given by all combinations of two independent variables with two levels each: \begin{itemize}
        \item Social Environment (Crowded, Uncrowded),
        \item Social Interaction Focus (Unfocused, Focused).
    \end{itemize}

    \subsubsection{Social Environment}
\begin{figure}[h!]
            \centering
            \subfloat[Crowded]{
            \includegraphics[height = 3.8cm, width = 4.15cm]{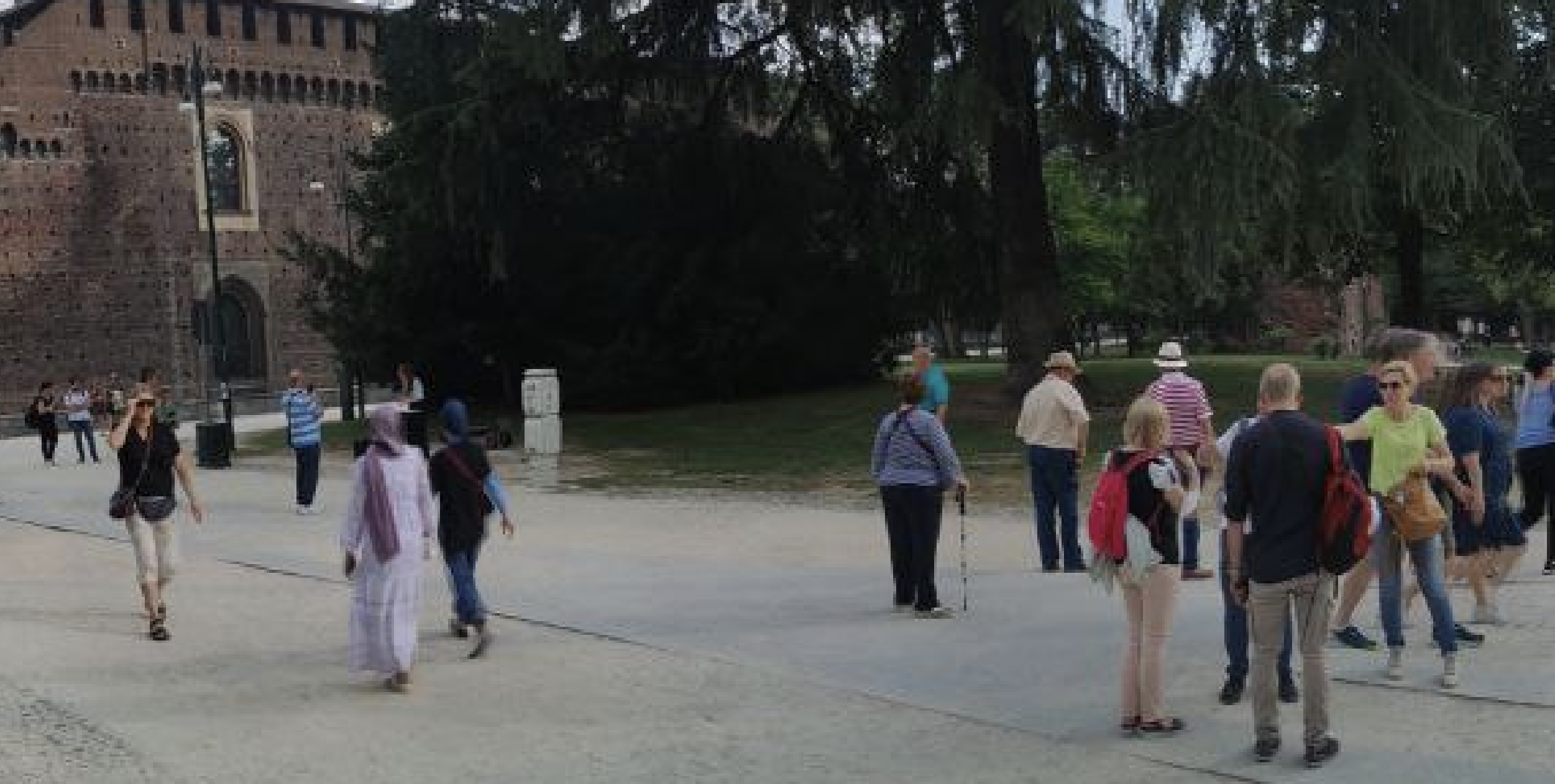}
            }
             \subfloat[Uncrowded]{
             
            \includegraphics[height = 3.8cm, width = 4.15cm]{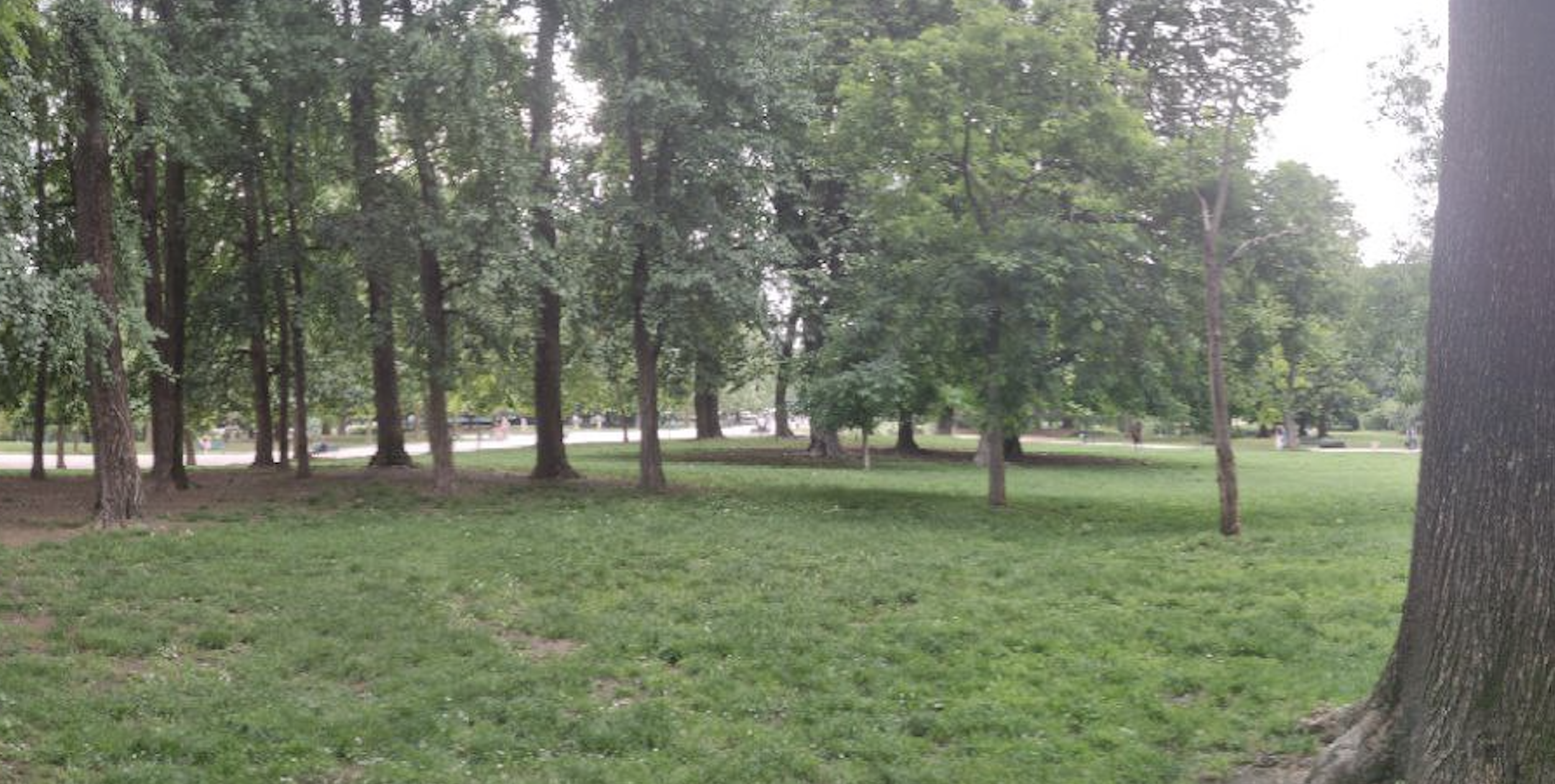}
            }
            
            \caption{Social Environments: a) Crowded - \enquote{Sforza's Castle, Milan, Italy}, b) Uncrowded - \enquote{Simplon Park, Milan, Italy}}
            \label{fig:environments}
        \end{figure}

    The environment of the game has to allow users to play in public. This means that players should feel like being in public, including crowded areas or landmarks of a city. At the same time, for the sake of the experiment, they should be able to play even in an uncrowded area. The choice to use AR for this application comes from the expressed environmental need. With AR, a clear connection with the physical world is maintained during the experiment, and at the same time, a digital layer is added to the experience. In consideration of this, two environmental conditions should be created: \textbf{\textit{(C) Crowded scenario, (U) Uncrowded scenario}} (see Fig. \ref{fig:environments}).
    A place in which the movement from Crowded to Uncrowded was easy and not too tiring was identified in places related to tourist landmarks close to city parks. With this vision, playing at the landmark is the Crowded scenario, while playing in the park is the Uncrowded one. Examples of such public spaces are probably present in most cities.
    
    \subsubsection{Focus}
    Focused social interaction is an interaction in which many players have a common goal. For the sake of the empirical study, it should be possible to have a setting in which social interactions between players are elicited and one in which this is avoided. Given the extreme flexibility of this game, this leads the research to have two other conditions to be considered:  \textbf{\textit{(0) Unfocused social interaction }}(or playing individually not sharing a common goal), \textbf{\textit{(1) Focused social interaction}} (sharing a common goal in a team).

    \subsection{Task Design}
    The task was to play a Scavenger Hunt (or Treasure Hunt) \footnote{en.wikipedia.org/w/index.php?title=Scavenger\_hunt\&oldid=1201190336}. The traditional concept of the game is that whoever organizes the game prepares some physical items to be found by the players. When playing, players must find those items, often hidden in some spots that are hard to find. In many versions of this game, clues are distributed to the players: just one clue may be distributed, and the next can be found only by finding the previous item. A competition layer was added by displaying a timer to users and declaring the time record for that particular setting before each experiment run. According to \cite{zagal2000model}, the choice to play a Scavenger Hunt fosters social interactions between players thanks to its particular rules: Cooperation is brought by the teamwork to find objects, The time score brings competition, and Meta-Gaming is present since the coordination on how to tackle the clue solution problem is completely left to the participants. It was decided to implement a Scavenger Hunt version in which clues were distributed sequentially to overview the players' behavior better while playing. Thus, at the start of the game, only the clue related to the first item is provided. When the player found it, the second was released, and so on. When playing in teams, each team member could collect the item for the whole team. However, this had to be a location-based game in an AR experience. Therefore, it was decided that the items to be found had to be digital, hidden in the physical space, with clues distributed digitally. This should be done with the same software deployed as a smartphone application to allow players to play with their phones. In this context, the Location-Based Augmented Reality technology serves as infrastructure to place items in specific places related to real physical spots (location-based feature) and to make the items appear overlaid over the information from the camera (Augmented Reality). According to Fonseca et al., \cite{fonseca2021requirements}, this game satisfies the framework's requirements for creating a good Location-Based Game (LBG). It establishes an underlying communication between players (Social Interaction), while the changing icon on the UI provokes Achievement and Reinforcement emotions. Lastly, it is intrinsically a Real-World Play. Following Fonseca's framework \cite{fonseca2021designing}, the players are conducting activities resembling Detective, Explorer, and Hunter.

    \begin{figure}[h!]
    \begin{tabular}{ccc}
        \centering
        \includegraphics[height = 5.5cm, width = 2.5cm]{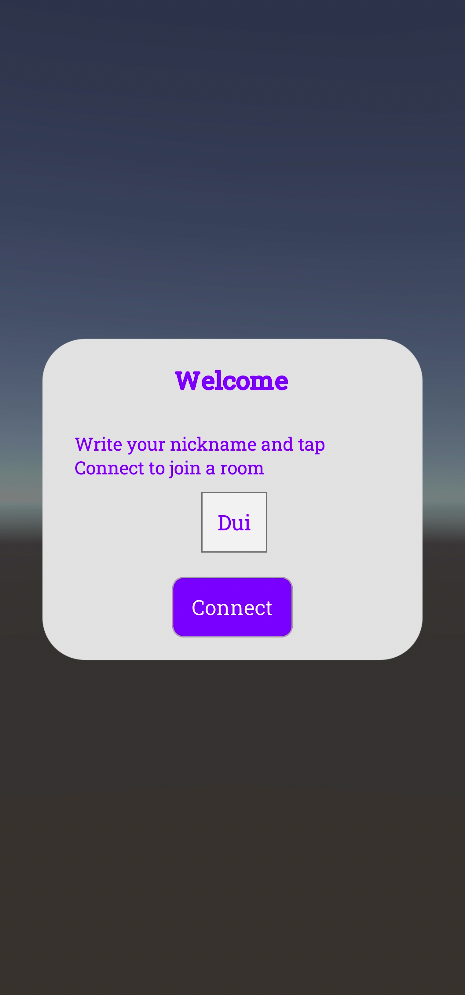} & \includegraphics[height = 5.5cm, width = 2.5cm]{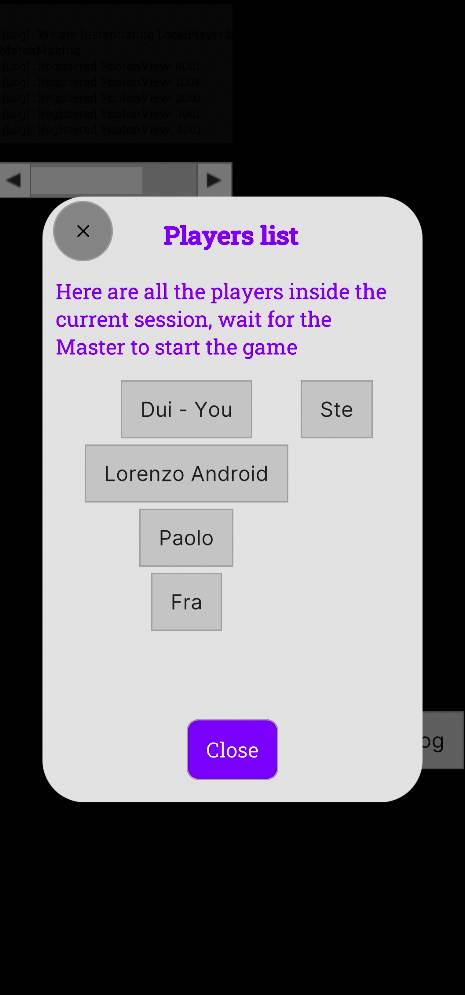} & \includegraphics[height = 5.5cm, width = 2.5cm]{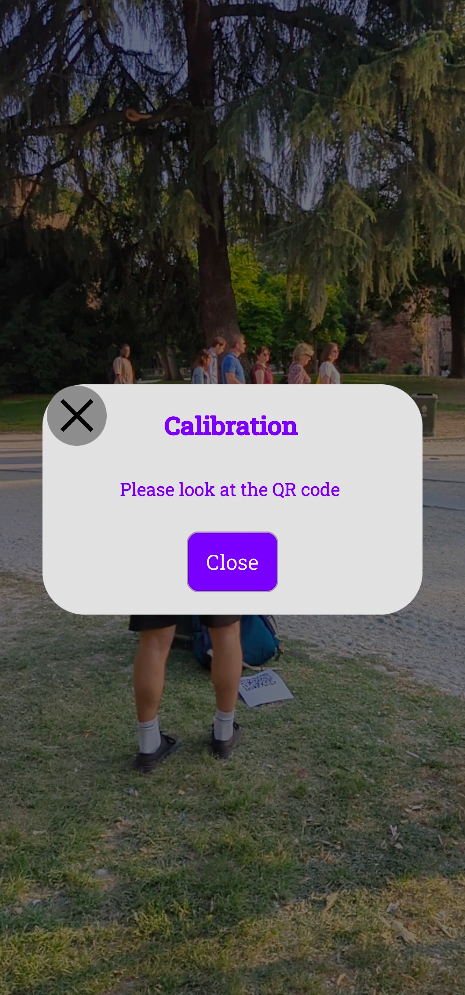}\\
        (a) & (b) & (c) 
    \end{tabular}    
    \caption[UI player - intro]{(a) Welcome and nickname input page. (b) Game lobby. (c) Scan QR code message}
    \label{fig:ui_1}
\end{figure}

\begin{figure}[h!]
    \centering
    \begin{tabular}{ccc}
        \includegraphics[height = 5.5cm, width = 2.5cm]{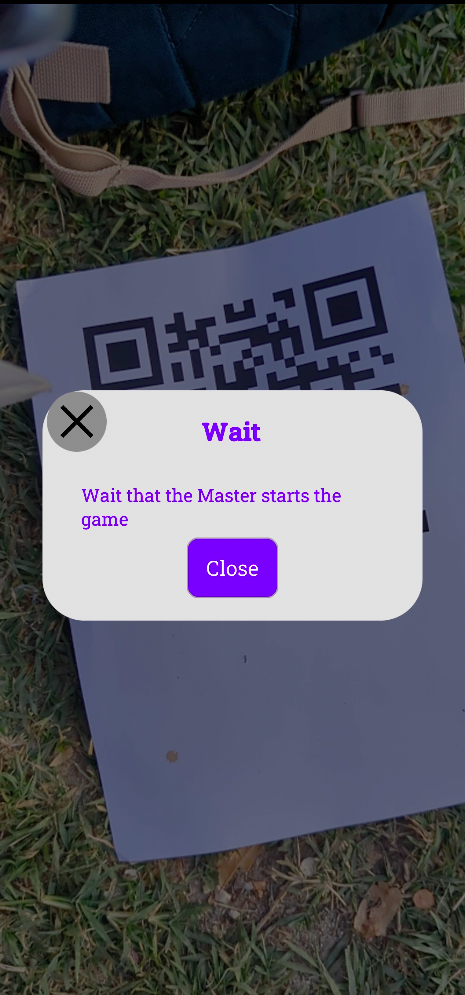} & \includegraphics[height = 5.5cm, width = 2.5cm]{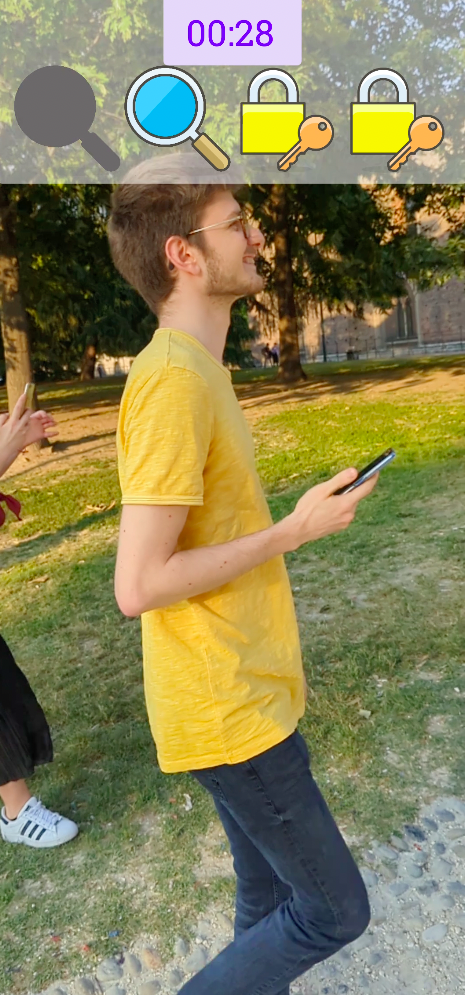} & \includegraphics[height = 5.5cm, width = 2.5cm]{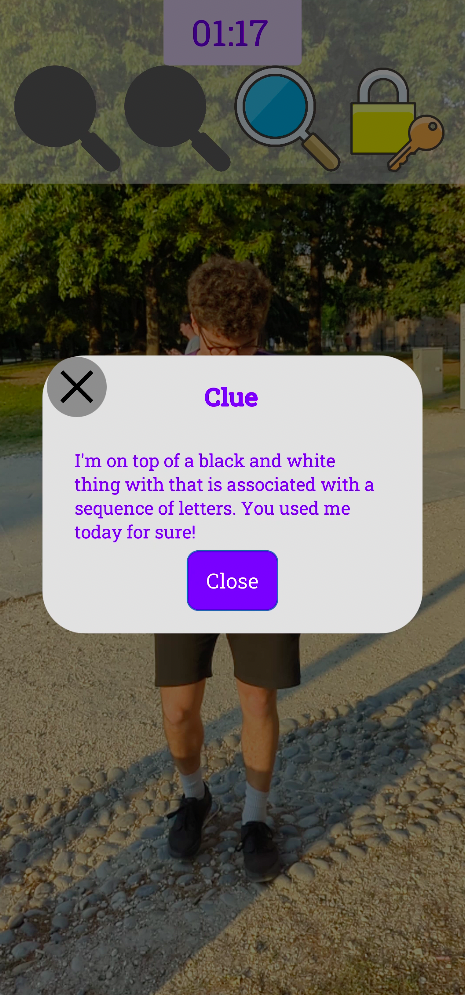}\\
        (a) & (b) & (c) 
    \end{tabular}    
    \caption[UI Player - Game Info UI]{(a) Successful QR code scan. (b) Timer, collected objects, remaining objects, unlocked objects. (c) Example of a clue}
    \label{fig:ui_2}
\end{figure}

\begin{figure}[h!]
    \centering
    \begin{tabular}{ccc}
        \includegraphics[height = 5.5cm, width = 2.5cm]{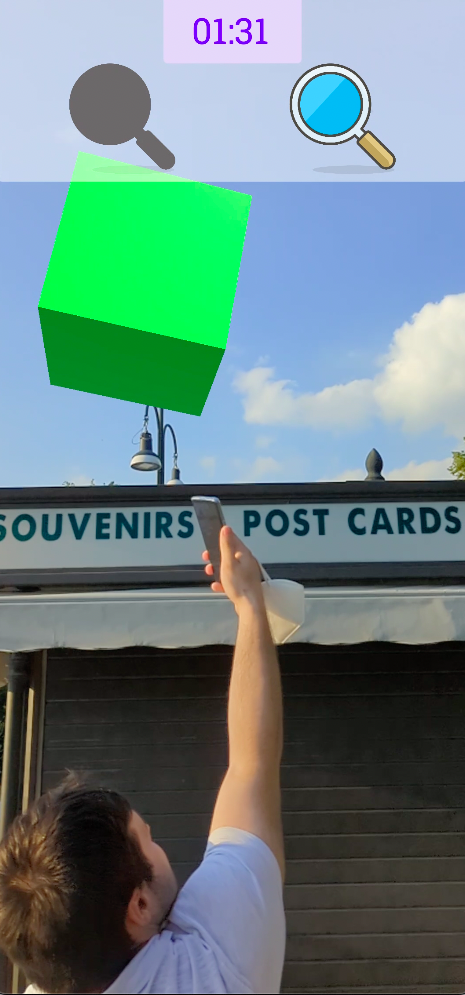} & \includegraphics[height = 5.5cm, width = 2.5cm]{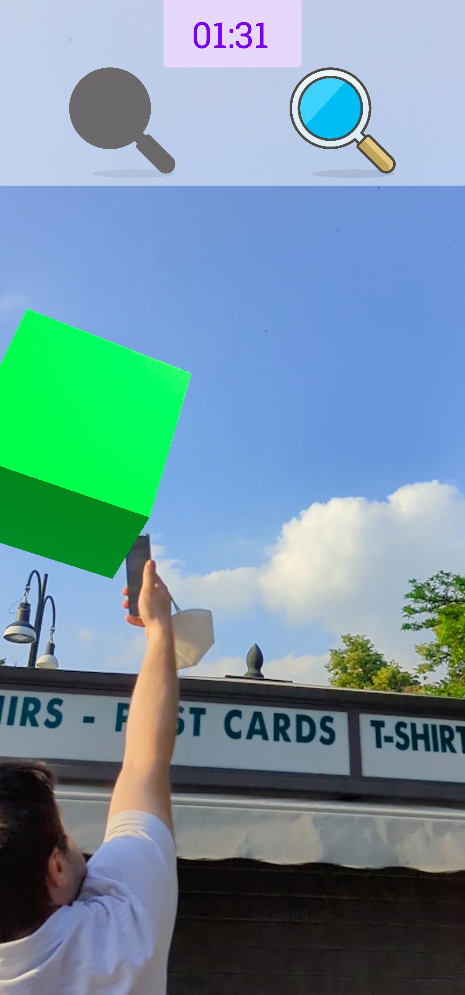} & \includegraphics[height = 5.5cm, width = 2.5cm]{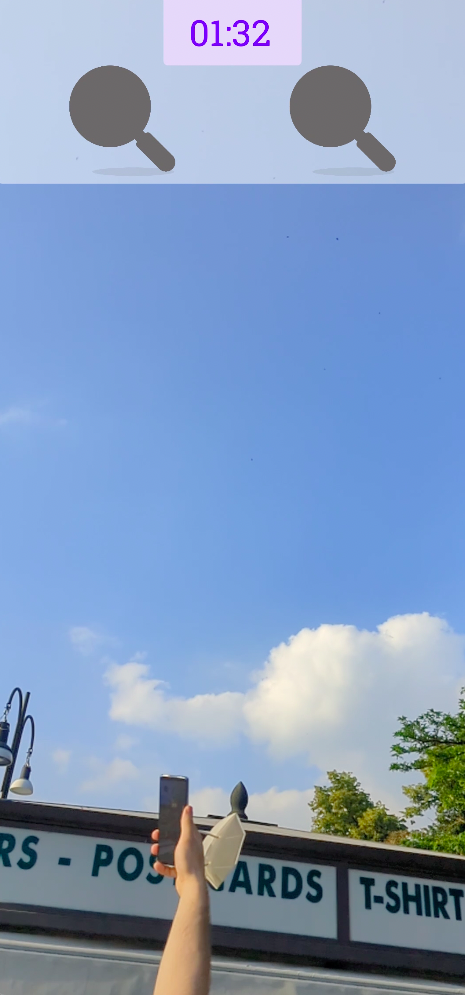}\\
        (a) & (b) & (c) 
    \end{tabular}    
    \caption[UI Player - Gameplay]{(a) Collectable object. (b) Bump the phone into the object to collect. (c) Collected object icon.}
    \label{fig:ui_3}
\end{figure}

    \subsection{Game Flow}
    
        \subsubsection{Moderator}
        The game starts by scanning a QR code placed in specific locations depending on the Crowded or Uncrowded setting. The moderator must select a specific configuration based on the condition (C0, C1, U0, U1), allowing them to place an item. Once the item is placed, a clue can be inputted in a banner that appears. This last step can be repeated for as many items as needed. When all the players join, the moderator starts the hunt.
        
        \subsubsection{Player}
        When the player starts the application, a banner pops up asking for a nickname to be inputted, and they can confirm by clicking the button \enquote{Connect} (see Fig. \ref{fig:ui_1}a). The player sees their nickname among those who already joined (see Fig. \ref{fig:ui_1}b). When the moderator starts the experience, the player's smartphone camera activates, and a banner asks to scan a QR code (see Fig. \ref{fig:ui_1}c). At this point, the player scans the previous QR code the moderator placed and waits for the hunt to start (see Fig. \ref{fig:ui_2}a). When the hunt begins, the game UI appears, the timer starts, and some items are assigned to be found (see Fig. \ref{fig:ui_2}b). The player can see the clue to reach the first item (see Fig. \ref{fig:ui_2}c), and when its location is reached, an object appears on top of what they saw from the camera (see Fig. \ref{fig:ui_3}a). The player hits the digital object with the phone as if it were in the physical space (see Fig. \ref{fig:ui_3}b). The UI shows that the item has been collected, and the next clue is available (see Fig. \ref{fig:ui_3}c). When all the items are collected, the game ends.

    \subsection{Procedure}
        Participants were invited to meet the moderator near a city landmark and public park. Initially, the participants were welcomed by a moderator and presented with an introduction to the study, including the game rules. After signing a consent form, the participants were given a pre-questionnaire about demographic information and their affinity for technology interaction. Then, a tutorial game is played to familiarise them with the UI and explain how it works. After this introductory part, the participant could start the condition communicated by the moderator. The participants were supposed to play all four conditions. After each condition, they had to stop and answer web-based questionnaires encompassing Short User Experience (UEQ-S) \cite{schrepp2017ueqs}, Networked Minds Social Presence Inventory (NMSPI) \cite{NMPI6743} and Social Acceptability \cite{koelle2018your}. Once all the conditions had been played and rated, the participants were asked to answer multiple-choice and open-ended questions in a final qualitative questionnaire. The complete duration of the experiment was between 45-60 minutes per participant.

\section{Results}
A repeated measure Analysis of Variance (ANOVA) was performed to determine statistically significant differences. An overview of all significant effects that will be explained in the following sections is given in Table \ref{tab:results}.

\begin{table}[h!]
  \centering
   \caption{Effects of Social Environments (Environment) and Social Interaction Focus (Focus) on Social Acceptability, Overall Social Presence, Perceived Psychological Engagement, Perceived Attentional Engagement, and Perceived Emotional Contagion.}
  \resizebox{\columnwidth}{!}{
\begin{tabular}{llcclrc}
\toprule
        Parameter & Effect & \( df_{n} \) & \( df_{d} \) & \( \hphantom{F}F \) & \( p\hphantom{p.} \) & \(\eta_{p}^{2} \) \\
\midrule
        Environment    & Social Acceptability     
        & 1  & 24  & 3.160  & .013 & .028  \\
        Focus    & Overall Social Presence  
        & 1  & 24  & 10.561  & .003  & .109  \\
        Focus    & Perceived Psychological Engagement  
        & 1  & 24  & 10.749  & .003  & .116  \\
        Focus    & Perceived Attentional Engagement
        & 1  & 24  & 27.268  & <.001  & .215  \\
        Focus    & Perceived Emotional Contagion   
        & 1  & 24  & 13.584  & .016 & .073  \\
        
\bottomrule
    \end{tabular}%
    }
  \label{tab:results}%
  \vspace{0em}
\end{table}%

\subsection{Social Environment}
\label{subsec: Social Environment}
The Social Environment independent variable has a statistically significant effect on the dependent variable of Social Acceptability (see Fig. \ref{fig:results}).
Results have shown that participants reported significantly higher average Social Acceptability for the Uncrowded environment conditions (U0: M= 2.46, SD= 0.93; U1: M= 2.60, SD= 0.88) compared to the Crowded environment conditions (C0: M= 2.20, SD= 1.21; C1: M= 2.10, SD= 1.39).
The effects of Social Environments on User Experience were not found to be significant. Nevertheless, it is interesting for later discussion to report the descriptive values (U0: M= 1.64, SD= 0.71; U1: M= 1.84, SD= 0.70, C0: M= 1.65, SD= 1.04; C1: M= 1.57, SD= 0.82).

\subsection{Social Interaction Focus}
\begin{figure*}[h]
\centering
\includegraphics[scale=0.4]{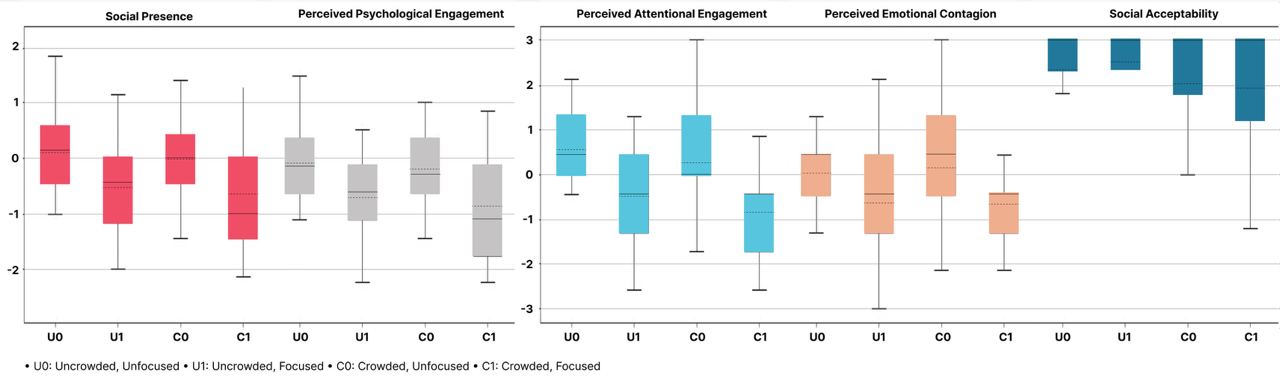}
\caption{Box plots of statistically significant results (-3...3). Whiskers indicate the first and third quartile, the continuous line indicates the median, and the dashed line indicates the mean values.}
\label{fig:results}
    
\end{figure*}

The independent variable Focus has a statistically significant effect on the dependent variables: Overall Social Presence, Perceived Psychological Engagement, Perceived Attentional Engagement, and Perceived Emotional (see Fig. \ref{fig:results}). 
The main effect of Focus found on the Overall Social Presence score shows higher average values for the Unfocused Interaction conditions (U0: M= 0.10, SD= 0.90; C0: M=0.01, SD= 0.95) compared to the Focused Interaction ones (U1: M=-0.53, SD = 0.84; C1: M= -0.65, SD= 0.97). When it comes to Perceived Psychological Engagement, participants have reported significantly higher average values for the Unfocused Interaction conditions (U0: M= 0.11, SD= 0.84; C0: M=0.01, SD= 0.96) compared to the Focused Interaction ones (U1: M=-0.53, SD= 0.86; C1: M= -0.68, SD= 1.03). Furthermore, results have shown significantly higher average Perceived Attentional Engagement scores for the Unfocused Interaction conditions (U0: M= 0.57, SD= 0.77; C0: M=0.26, SD= 1.14) compared to the Focused Interaction ones (U1: M=-0.50, SD= 1.07; C1: M= -0.86, SD= 1.13). For what concerns Perceived Emotional Contagion, the results have shown significantly higher average scores for the Unfocused Interaction conditions (U0: M= 0.02, SD = 1.31; C0: M=0.15, SD= 1.41) compared to the Focused Interaction ones (U1: M=-0.63, SD= 1.24; C1: M= -0.67, SD= 1.27). Focus had no significant effect on User Experience (see Subsec. \ref{subsec: Social Environment} for UX descriptive values).

\subsection{Qualitative Results}
The final questionnaire included three open questions about general impressions of the testing procedure and the game. Some users were bothered by repeatedly filling in questionnaires regarding the testing process. Other feedback was about how teams were formed, saying it would have been nice to try the game out with total strangers. Finally, a user suggested competitive sessions, with more teams competing against each other. Regarding the game, some participants suggested better position management and including other objects rather than just green boxes. Some positive feedback was \enquote{It's hard to imagine a clearer way to connect gamer and game}. Thus, this player felt like being inside a digital game in which they were the main character. In terms of future scenarios, many people suggested the game for children's events like birthdays or scavenger hunts organized by the school. Some comments were about company events to improve teamwork skills, others as a game to play with friends with physical prizes (like snacks or drinks) for picnics. Finally, many users mentioned the touristic scenarios where one could find hidden spots or landmarks brought there by riddles. Then, according to users, it could be used in zoos or museums. 

\section{Discussion}

\subsection{Social Presence}
Results show higher social presence scores in the conditions in which they are playing alone and lower social presence scores in the conditions in which they are in teams. This means that users who were playing in teams were more cautious about their teammates' thoughts, feelings, and actions, while when they were playing individually, they did not feel much influenced by the others and felt like others' attention was addressed to them. This result is supported by statistical analysis as well. Two-way repeated measures ANOVA results show significant statistical differences between the means of many (but not all) Social Presence-related dimensions. Each of these is affected by the social interaction Focus. The overall score and the Perceived Psychological Engagement, Attentional Engagement, and Emotional Contagion are affected by the focus of social interaction.

Looking at the values of this piece of data, it is surprising how the absolute values are small. The mean values are all in the range (-0.86, 0.57), and even quartiles rarely lie in the absolute values (2,3). An interesting case is the one coming from Perceived Attentional Engagement. This dimension gives lower scores if users feel that their attention was addressed to other teammates, and higher scores result from users self-reporting that they felt that teammates were addressing their attention to them. This is when the difference between the means is the highest. This shows how strongly users were convinced that their attention was directed to others when playing in teams and how they felt like their teammates' attention was addressed to them when they played alone.

\subsection{Social Acceptability}
ANOVA results show a significant difference in the means driven by the Social Context. Statistically, users regarded the Uncrowded setting as more acceptable than the Crowded one. This result follows what intuition could state: people should regard a more intimate public context as more socially acceptable, while a public context in which many people are passing around looking at you should be considered less socially acceptable. Nevertheless, even the more intensive public context setting is socially accepted, showing high scores in the self-reported questionnaire.

\subsection{User Experience}
When analyzing the User Experience results, Schrepp et al. \cite{schrepp2017construction} benchmark was used to compare the results obtained. UEQ-S overall means lay between 1.57 and 1.84, depending on the conditions. According to the benchmark comparison, these mean values correspond to an experience marked around the distinction between the average mean related to Good and Excellent. Therefore, the overall score of UEQ-S can tell that users had a Good to Excellent experience with the game. Considering Hedonic and Pragmatic qualities, it is visible how all the means related to the Hedonic quality are higher than the Pragmatic one. According to the benchmark, all hedonic qualities lie in the excellent experience area, while most pragmatic qualities are good. This suggests that the most preferred quality from the user was not related to the tasks but more to the leisure or fun while playing. Qualitative results align with this vision since the negative feedback was not about the boredom of the experiment but about the complexity of tasks or an improvement in the collectible position management and their relative riddles. The mean values show how condition U1 gave higher results in Overall Pragmatic and Hedonic qualities. This condition is relative to the Uncrowded public context and team play. According to UEQ-S results, this configuration is the most preferred one. Thus, it is possible to conclude by simple mean comparison from different dependent variables that the Social Environment and the social interaction Focus affect UX qualities, and users prefer to play in an Uncrowded setting and teams or with a Focus on Social Interaction between them. However, ANOVA results show no statistical relevance to the difference in the means between the conditions. Therefore, no further conclusion can be given from statistics. Thus, according to the statistical analysis that has been made, UX dimensions are not affected by the variation of public space or social interaction focus conditions.

\section{Conclusion}

In this research work, the Social Environment and Social Interaction Focus were the conditions of an empirical study conducted in a public space. This work aimed to understand how these conditions could affect User Experience and Social Acceptability in a game based on Augmented Reality. Particular attention was brought to Social Presence, which overlaps with User Experience and Social Acceptability. An AR application was specifically designed for this experiment. The game was a Scavenger Hunt-based game: players had to solve riddles to find digital items only visible if users were close enough to them in the physical public space. Results were built by quantitative analysis of data from different types of questionnaires filled out by the participants. This analysis included descriptive statistics and two-way repeated measures analysis of variances (ANOVA).
Furthermore, qualitative data coming from observation was considered. Regarding Social environments, it has been found that there is statistical evidence of users' self-reported Social Acceptability differences between the settings. Results report generally high scores as if users had no problems with the public environment.
Nevertheless, Uncrowded public settings were preferred, in particular, the setting of playing in teams. This aspect, which comes from quantitative data analysis, is also supported by qualitative analysis. Users hardly complained about the public setting and never seemed embarrassed while playing the game in front of strangers and a public space.
Moreover, they looked forward to playing those team games once they understood the dynamics. Social Interaction Focus was statistically important in Social Presence. Users cared more about people experiencing co-presence with them in unfocused settings than focused ones. Statistical evidence supports these findings. When playing in teams, users thought more about their role in the game and were more involved in bringing their contribution to the team. Playing individually, on the other hand, they were not thinking anymore about their role inside the game but more about how their behavior was seen by their teammates and passers-by who were looking at the player playing the game.

\subsection{Limitations}

Some limitations have been identified with this work.
A within-subject design applied in public has the risk of reporting biased results from users' backgrounds. In other words, user recruitment could have been more detailed. For example, users could have been foreigners in the public space where the empirical study is done. Moreover, more data could have been gathered from users to understand different affective dimensions. The full UEQ questionnaire could have been used rather than the short version, together with other surveys such as Self-Assessment Manikin and a between-subject design.

\subsection{Future Work}

More empirical research has to be done in this area to understand how users interact with technologies and other people while using immersive technologies.
Future studies should consider different technologies such as VR, MR, or AR paradigms. More dependent variables should be considered, such as all the UEQ dimensions, and the study design should consider other types, such as the between-subject design.
Finally, these future studies should not only involve a larger number of participants and a more heterogeneous group in terms of nationality but also take into account precisely their profiles, such as their ATI profile and their education.

\bibliographystyle{IEEEtran}
\bibliography{IEEEabrv, main}

\end{document}